\documentclass[11pt]{article}

\usepackage{amsfonts,amsmath,amsxtra}
\usepackage{amssymb,amscd}
\usepackage[all]{xy}

\catcode`\@=11
\def\marginnote#1{}
\newcount\hour
\newcount\minute
\newtoks\amorpm
\hour=\time\divide\hour by60
\minute=\time{\multiply\hour by60 \global\advance\minute by-\hour}
\edef\standardtime{{\ifnum\hour<12 \global\amorpm={am}%
     \else\global\amorpm={pm}\advance\hour by-12 \fi
     \ifnum\hour=0 \hour=12 \fi
   \number\hour:\ifnum\minute<10 0\fi\number\minute\the\amorpm}}
\edef\militarytime{\number\hour:\ifnum\minute<10 0\fi\number\minute}
\def\draftlabel#1{{\@bsphack\if@filesw {\let\thepage\relax
\xdef\@gtempa{\write\@auxout{\string
   \newlabel{#1}{{\@currentlabel}{\thepage}}}}}\@gtempa
\if@nobreak \ifvmode\nobreak\fi\fi\fi\@esphack}
     \gdef\@eqnlabel{#1}}
\def\@eqnlabel{}
\def\@vacuum{}
\def\draftmarginnote#1{\marginpar{\raggedright\scriptsize\tt#1}}
\def\draft{\oddsidemargin -0.1truein
     \def\@oddfoot{\sl preliminary draft \hfil
     \rm\thepage\hfil\sl\today\quad\militarytime}
     \let\@evenfoot\@oddfoot \overfullrule 3pt
     \let\label=\draftlabel
     \let\marginnote=\draftmarginnote
\def\@eqnnum{{\rm (\theequation)}
\rlap{\kern\marginparsep\tt\@eqnlabel}%
\global\let\@eqnlabel\@vacuum}  }
\def\numberbysection{\@addtoreset{equation}{section}
     \def\theequation{\thesection.\arabic{equation}}}
\numberbysection

\renewcommand{\theequation}{\thesection.\arabic{equation}}
\makeatletter
\newdimen\normalarrayskip            
\newdimen\minarrayskip               
\normalarrayskip\baselineskip
\minarrayskip\jot
\newif\ifold             \oldtrue            \def\new{\oldfalse}
\def\arraymode{\ifold\relax\else\displaystyle\fi}
\def\eqnumphantom{\phantom{(\theequation)}} 
\def\@arrayskip{\ifold\baselineskip\z@\lineskip\z@
  \else
  \baselineskip\minarrayskip\lineskip1\baselineskip\fi}
\def\@arrayclassz{\ifcase \@lastchclass \@acolampacol \or
\@ampacol \or \or \or \@addamp \or
\@acolampacol \or \@firstampfalse \@acol \fi
\edef\@preamble{\@preamble
\ifcase \@chnum
  \hfil$\relax\arraymode\@sharp$\hfil
  \or $\relax\arraymode\@sharp$\hfil
  \or \hfil$\relax\arraymode\@sharp$\fi}}
\def\@array[#1]#2{\setbox\@arstrutbox=\hbox{\vrule
  height\arraystretch \ht\strutbox
  depth\arraystretch \dp\strutbox
width\z@}\@mkpream{#2}\edef\@preamble{\halign \noexpand\@halignto
\bgroup \tabskip\z@ \@arstrut \@preamble \tabskip\z@ \cr}%
\let\@startpbox\@@startpbox \let\@endpbox\@@endpbox
\if #1t\vtop \else \if#1b\vbox \else \vcenter \fi\fi
\bgroup \let\par\relax
\let\@sharp##\let\protect\relax
\@arrayskip\@preamble}
\def\eqnarray{\stepcounter{equation}%
           \let\@currentlabel=\theequation
           \global\@eqnswtrue
           \global\@eqcnt\z@
           \tabskip\@centering              
           \let\\=\@eqncr
           $$%
         \halign to \displaywidth  \bgroup
          \eqnumphantom \@eqnsel
   \hskip\@centering                               
 $\displaystyle  \tabskip\z@ {##}$%
 &\global\@eqcnt\@ne \hskip 2\arraycolsep
      $ \displaystyle  \arraymode{##}$\hfil
 &\global\@eqcnt\tw@ \hskip 2\arraycolsep
      $\displaystyle\tabskip\z@{##}$\hfil
      \tabskip\@centering
 &{##}\tabskip\z@\cr}
\makeatother

\newcounter{mo}

\newcounter{bk}

\newcommand{\tr}{{\rm tr}}

\newcommand{\om}{\omega}

\newcommand{\de}{\delta}
\newcommand{\al}{\alpha}
\newcommand{\te}{\theta}

\newcommand{\be}{\beta}

\newcommand{\ka}{\kappa}

\newcommand{\si}{\sigma}

\def\bea{\begin{eqnarray}\new\begin{array}{cc}}
\def\ee{\end{array}\end{eqnarray}}

\newcommand{\beq}[1]{\begin{equation}\label{#1}}
\newcommand{\eq}{\end{equation}}
\newcommand{\beqn}[1]{\begin{small} \begin{eqnarray}\label{#1}}
\newcommand{\eqn}{\end{eqnarray} \end{small}}
\newcommand{\p}{\partial}
\def\sq2{\sqrt{2}}
\newcommand{\di}{{\rm diag}}

\newcommand{\SL}{{\rm SL}(2,{\mathbb C})}

\def\sl2{{\rm sl}(2, {\mathbb C})}

\def\f1#1{\frac{1}{#1}}

\def\mC{{\mathbb C}}
\def\mZ{{\mathbb Z}}
\def\mR{{\mathbb R}}

\def\frak{\mathfrak}

\def\gg{{\frak g}}

\def\gm{{\frak m}}

\def\gk{{\frak k}}

\def\gu{{\frak u}}

\def\bfX{{\bf X}}
\def\bfY{{\bf Y}}

\def\bfz{{\bf z}}

\def\clF{\mathcal{F}}

\def\clI{\mathcal{I}}

\def\clM{\mathcal{M}}

\def\clX{\mathcal{X}}

\def\clZ{\mathcal{Z}}

\def\bag2{{\bf g_2}}
\def\bas8{{\bf so(8)}}

\def\sr2{\sqrt{2}}

\def\f1#1{\frac{1}{#1}}

\begin{document}


 \begin{flushright}
 ITEP-TH-35/24\\
IITP-TH-30/24
 \end{flushright}
\vspace{3mm}

 \begin{center}
{\Large\bf Families of Kuramoto models }
\\ \vspace{4mm}
{\Large\bf and bounded symmetric domains}\\
 \vspace{7mm}
  {\large M. Olshanetsky}\\
   \vspace{3mm}
  {\rm
National Research Centre "Kurchatov Institute",\\
  Academician Kurchatov square, 1, Moscow, 123182, Russia
}\\
   \vspace{1mm} {\rm Institute for Information Transmission Problems RAS (Kharkevich Institute),
 \\  Bolshoy Karetny per. 19, Moscow, 127994,  Russia}\\
\vspace{2mm}
 {\footnotesize Email
 mikhaail.olshanetsky@gmail.com
 }
 \end{center}


 \begin{abstract}
 We define the families of Kuramoto models (KM) related to
 bounded symmetric domains.
 The families include the Lohe unitary model and
 the spherical models as special cases. Our approach is based on the construction proposed by Watanabe and
 Strogats WS. We replace the Poincare disc and its $S^1$
 boundary in the WS construction on the bounded symmetric domains and on the its Bergman-Shilov (BS) boundaries.
  In Cartan classifications there are four classical
 domains of types I-IV. Here we consider the domains of types I,II
 and III.
  For a fixed domain there is a decreasing chain
 of the BS boundaries components. This leads to the KM families we described here.
 \end{abstract}

\section{Intruduction}

The Kuramoto model \cite{Ku} (KM) describes synchronisation behaviour in
 physics, biology, chemistry and the social sciences. The model is defined by
 $N$ pairwise interacting particles located on a circle (oscillators with phases $\te^J$) governed by
 the system of equations
\beq{sk0}
\dot{\te}^J=
\om+\frac{2\ka}N\sum_{I=1}^N\sin(\te^J-\te^I)\,,~~(J=1,\ldots,N)\,,
\eq
 Much work has been devoted to the study of this model and its numerous applications
 (see review \cite{ABV}).
 An important insight into the structure of KM was provided by
 Watanabe and Strogats (WS) \cite{WS}. They considered the complexification
 of the phases $z^J=e^{\imath\te^J}$ and continued the flows (\ref{sk0}) from the circle $S^1$ to the interior disc
$D=\{|z^J|<1|\}$. The disc, equipped  with the hyperbolic geometry, becomes
a Poincar\'e model of the Lobachevsky plane. The Lobachevsky metric is invariant under the
action of the group $G_M$ of M\"{o}bius transformations. The Kuramoto flows (\ref{sk0})
on $D$
are generated by the Lie algebra action of $G_M$. The flows can be matched to a  flow on the group $G_M$. This is how Watanabe and Strogats  discovered a hidden symmetry in the Kuramoto model. This approach is described in detail in \cite{CEM,MMS}.

The aim of this paper is to generalise the WS construction to flows on some multidimensional
manifolds. These are
 bounded symmetric domains (BSD) and some of their special boundaries (the Bergman-Shilov (BS) boundaries). In special cases the BSD construction is reduced to the WS construction or
to its previously proposed generalisations.
First, we replace the group $G_M$ on the quasi-unitary group $G=$SU$(1,1)$.
\footnote{ In contrast to SU$(1,1)$ the group $G_M$ is not unimodular, although the both groups
act by the  M\"{o}bius transformations and
preserve the disc.}
Its actions also generates the Kuramoto flows.

The bounded symmetric domains are classified by E.Cartan \cite{He}.
There are four classical series and two exceptional cases (see the Table in the Appendix).
Here we  confine ourselves to considering
 the first three classical types of BSD.

Let us briefly illustrate our scheme for the Type I BSD.
Define  the space of complex rectangular
matrices  $Z_{mn}$ satisfying the inequality (\ref{ft}). They form the BSD of type I $D^I_{mn}$. Let  $G=$SU$(m,n)$ be  the group
 of complex unimodular matrices preserving the Lorentzian
form
\beq{sum}
{\rm SU}(m,n)=\{h\,|\, hsh^\dag=s\,,~s=(\underbrace{-1,\ldots,-1}_m,\underbrace{1,\ldots,1}_n)
\,,~ \det\,h=1\,,~h^\dag=\bar h^T\}\,.
\eq
 The group $G$ acts on $Z_{mn}$ by the matrix M\"{o}bius transformation.
 The domain  $D^I_{mn}$ is non-compact.
 Assume that $m\geq n$. The boundary of  $D^I_{mn}$  has a $n$ components.
 The components are homogeneous spaces with respect to the  M\"{o}bius transformations.
 All but one of the components  are open. The open components
 are equivalent to BSD $D^I_{m-t,n-t}$, for $t=1,\ldots,n-1$.
  The closed component is the analogue of the circle $S^1$ in the WS construction. It is called the BS boundary.
  This boundary arises in the theory of functions of many complex variables as part of the topological boundary of a bounded domain. The holomorphic functions inside the domain are completely determined by their values on the BS boundary.
For the type I domains the BS boundary  is the complex Stiefel manifold $St^\mC_{mn}$ (\ref{sm}). If $m=n$
the Stiefel manifold  is the manifold of unitary matrices $St^\mC_{m,m}\sim {\rm U}(m)$.
For $n=1$ the Stiefel manifold  is the odd-dimensional sphere $St^\mC_{m,1}\sim S^{2m-1}$.
The group $G$ continues to act on $St^\mC_{mn}$. The action of its Lie algebra generates
a flow in the form of the matrix Riccati differential equation.
It allows one to define the matrix Kuramoto
flows on this component.
The open components have their own BS boundaries $St^\mC_{m-t,n-t}$ of smaller dimension.
 The subgroups ${\rm SU}(m-t,n-t)$ act on them. This gives us  the Kuramoto
flows on $St^\mC_{m-tn-t}$. In this way we get the type I  Kuramoto families, which are related to the
type I bounded domains. In particular the family related to BSD $D^I_{mm}$ is the Lohe unitary
family $U(m),U(m-1),\ldots U(1)$  \cite{Lo,NA}.
It ends with the standard Kuramoto model. The family related to the domain $D^I_{m1}$
contains only one term. It is the odd-dimensional sphere Kuramoto model \cite{Cr,LMS,Ta}.

The BSD generalisation of the WS construction for the Type I models is as follows
\begin{center}
\begin{tabular}{|c|c|}
\hline
WS construction & BSD construction\\
  \hline
  Unit disk $|z|<1$& BSD \\
 Boundary $|z|=1$ & BS boundary=Stiefel manifold  \\
 Symmetry group $G_M$ & $G=$SU$(m,n)$  \\
  M\"{o}bius transformation & matrix M\"{o}bius transformation \\
  \hline
\end{tabular}
\end{center}

In the next section we describe the WS construction  in terms that we use in the general case.
The detailed construction of the Type I Kuramoto model is given in Section 3.
The construction of the Type II and III Kuramoto models is outlined in Sections 4 and 5.
It is largely the same as that of the
Type I.
The classification of BSDs
 is given at the end of the article.
 Information about BSDs and their boundaries can be found in \cite{Cl,He,Low,PS,Wo}.

  Here we consider
 only first three types of the classical domains. The remaining cases will be discussed in a subsequent publication.


\section{Preliminaries about SU(1,1) and Poincar\'{e} disk}

Let $g\in\SL$,
$g=\left(
     \begin{array}{cc}
       A & B \\
       C &  D\\
     \end{array}
   \right)
   $, $\det\,g=1$, $(A,B,C,D)\in\mC$. Consider its two real subgroups
   \beq{1}
   {\rm SU}(2)=\left\{u=\left(
     \begin{array}{cc}
       A & B \\
       -\bar B &  \bar A\\
     \end{array}
   \right)\,,~|A|^2+|B|^2=1\right\}\,,~~{\rm  unitary~group}\,,
   \eq
   \beq{2}
   {\rm SU}(1,1)=\left\{h=\left(
     \begin{array}{cc}
       A & B \\
       \bar B &  \bar A\\
     \end{array}
   \right)
   \,,~|A|^2-|B|^2=1\right\}
   \,,~~{\rm  pseudo-unitary~group}\,.
   \eq

The group SU$(1,1)$ preserves the Lorentz form $s=\di(-1,1)$
$$
{\rm SU}(1,1)=\{g\,|\,gsg^\dag=s\}\,,~~g^\dag={\bar g}^T\,.
$$

The Lie algebra Lie(SU$(1,1))=\gg$ has the matrix representation
\beq{ls}
\gg=\left\{\left(
     \begin{array}{cc}
       a & b \\
       \bar b &  -a
     \end{array}
   \right)
   \,,~\Re e\,a=0\right\}\,.
\eq

 Let $K$ be the maximal compact subgroup of SU$(1,1)$. It can be represented by the diagonal matrices
   \beq{k0}
   K={\rm S(U}(1)\times{\rm U}(1)=\{\di(e^{\imath\te},e^{-\imath\te})\}\,.
   \eq

   The quotient
   \beq{x0}
   X={\rm SU}(1,1)/{\rm S(U}(1)\times{\rm U}(1))
   \eq
  is the Poincare disc.

\noindent
\emph{Borel embedding}

Let $P$ be the parabolic subgroup of $\SL$
$$
P=\left\{\left(
     \begin{array}{cc}
       A & 0 \\
       C & A^{-1}\\
     \end{array}
   \right)\right\}\,.
$$
  The Iwasawa decomposition  for $\SL$ means that any $g\in \SL$ can be represented
   in the form
   \beq{iw}
   g=up\,,~u\in {\rm SU}(2)\,,~p\in P\,,~
   {\rm SU}(2)\cap P=K\,,
   \eq
   This decomposition can be checked directly using the representations
   of SU$(2)$ and $P$.
   The right action of the parabolic subgroup leaves invariant the relation
   $z=b/d$. Since $|b|^2/|d|^2=|b|^2/|a|^2$ and $|a|^2+|b|^2=1$
   $z$ belongs to the complex plane $\mC$ $(d\neq 0)\sim (z\neq\infty)$.

   It means that the
   quotient $\SL /P$ for $g\in\SL$ with $d\neq 0$ is the complex plane $\mC$.
   the  matrices
From (\ref{iw}) follows two isomorphic description of the sphere $S^2\sim\mC P^1$
\beq{be}
\mC P^1=\SL/P\sim  {\rm SU}(2)/K=X_c\,.
\eq
This compact space is the Cartan dual to the Poincar\'{e} disc $X$ (\ref{x0}).

   Now replace in (\ref{iw})  SU$(2)$ on SU$(1,1)$.
   \beq{iw1}
  \bfY=\{ g=up\,,~u\in {\rm SU}(1,1)\,,~p\in P\}\,,~
   {\rm SU}(1,1)\cap P=K=\{\di(e^{\imath\te},e^{-\imath\te})\}\,.
   \eq
Thus
\beq{be1}
\bfY/P\sim X={\rm SU}(1,1)/K
\eq

Because in this case  $|a|^2-|b|^2=1$ we have for $z=b/a$ $|z|^2=|b|^2/|a|^2<1$.
  Thereby the noncompact symmetric space $X$ is embedded as an open
disk $|z|<1$    in  the compact dual space $X\subset\mC P^1$
and the space $\bfY$
is only inclusion  $\bfY\subset\SL$ as an open subset.
\beq{be0}
X\subset X^c\,.
\eq
It is the Borel embedding of the noncompact space $X$ (\ref{x0})  as an open subset in the compact
space $X_c=\mC P^1$ (\ref{be}).

\noindent
\emph{Harish-Chandra embedding.}\\
Here we clarify the representation of $X$ as the unite disk
using the Gauss decomposition of $\SL$.
Let
\beq{HC}
\gm^+=
\left\{\left(
             \begin{array}{cc}
               0 & z \\
               0 & 0 \\
             \end{array}
           \right)\,,~~z\in\mC
\right\}\,,
~~
 M^+=\exp(\gm^+)=Id_{2}+\left(
             \begin{array}{cc}
               0 & z \\
               0 & 0 \\
             \end{array}
           \right)\,.
          \eq
Let $K^\mC=\di(k,k^{-1})$, $k\in\mC^*$. Consider the product
\beq{x}
{\bf X}=M^+K^\mC M^-=M^+P\,.
\eq
 Any $g\in\SL$ has the form (\ref{x})
except $d=0$. So $\bfX$ is a dense and open in $\SL$
 (the Gauss decomposition)
\beq{gd1}
\SL=\overline{\bf X}=\overline{M^+P}\,.
\eq
 For $g\in$SU$(1,1)$
$d=\bar a$ and $|a|^2\geq 1$. Therefore, SU$(1,1)$ can be represented in
the form (\ref{x}).
We show that for  SU$(1,1)$ the element $M^+$ in the Gauss decomposition (\ref{x})  is not arbitrary, but $|z|<1$.
 Any element $g\in$SU$(1,1)$ has the Cartan decomposition
\beq{cr1}
g=k_1hk_2\,,~~k_{1,2}=\di(e^{2\imath\pi\te_{1,2}},e^{-2\imath\pi\te_{1,2}})\,,
\eq
where
$$
h=h(t)=\exp
\left(
    \begin{array}{cc}
     0  & t  \\
      t& 0 \\
    \end{array}
  \right)
=\left(
    \begin{array}{cc}
     \cosh t  & \sinh t \\
      \sinh t & \cosh t \\
    \end{array}
  \right)\,,~~(t-{\rm real})\,.
$$
The matrix $k_1h$ represents an element of the quotient $X$ (\ref{x0}).
The representation (\ref{cr1}) can be transformed in the Gauss form (\ref{x}). First consider
the Gauss decomposition of $h$
$$
h=\left(
  \begin{array}{cc}
    1 & \tanh t\\
    0 & 1 \\
  \end{array}
\right)
\left(
  \begin{array}{cc}
   1/ \cosh t & 0\\
    0 & \cosh t   \\
  \end{array}
\right)
\left(
  \begin{array}{cc}
    1 &0\\
    \tanh t & 1 \\
  \end{array}
\right)\,.
$$
Then
\beq{cr}
k_1h=e^{2\imath\pi\te_1}
\left(
  \begin{array}{cc}
    1 & \tanh t\\
    0 & 1 \\
  \end{array}
\right)P=\left(
  \begin{array}{cc}
    1 & z\\
    0 & 1 \\
  \end{array}
\right)P\,,~~(\te_1\in[0,1))\,,
\eq
Thus $z=e^{2\imath\pi\te_1}\tanh t$ and $|z|<1$. In this way the space $X$ (\ref{x0})
is identified with the unit disk (the Poincar\'{e} disk model of the Lobachevsky plane)
\beq{ld}
D=\{z\in\mC P^1\,|\,|z|<1\}\sim X~(\ref{x0})\,.
\eq
For further purposes we rewrite it as follows.
Let
$$
Z=\left(
    \begin{array}{cc}
    0 & z \\
    0 & 0 \\
    \end{array}
  \right)\,.
$$
 The condition $|z|<1$ can be rewritten
\beq{ft0}
 D=\{z\in \mC\,|\,Id_2   -ZZ^\dag>0\}\,,~~(Z^\dag=\bar Z^T)\,,
  \eq
where the inequality $>$ means that the eigenvalues of the matrix are positive.
 It follows from (\ref{cr}) that we defined the map
 \beq{hc0}
 \xi\,:\, D\to \mC P^1\,.
 \eq
 It is the Harish-Chandra map.  Since  $K\subset P$, the nilpotent matrix
 $\left(
  \begin{array}{cc}
    1 & z\\
    0 & 1 \\
  \end{array}
\right)$ in (\ref{cr}) represents an element of the quotient $X_c$ (\ref{be}). Thereby
 the image of $D$ is the quotient $X$ and $X\subset\mC P^1=X_c$.

The boundary of $D$ in $\mC P^1$ is the circle $\p D=S^1$. In terms of the matrix description
$$
\p D=\{\det\,(Id_2   -ZZ^\dag)=0\}
$$
The group SU$(1,1)$ (\ref{2}) acts transitively on $D$ as well as on its boundary
$\p D$ by the M\"{o}bius transformation
\beq{mu1}
h(z)=\frac{Az+B}{\bar Bz+\bar A}\,,~~h\in\,{\rm SU}(1,1)\,.
\eq
Note that the center $\clI=\{\pm Id_2\}$ of ${\rm SU}(1,1)$ does not act on $D$.
So the group we need is the quotient ${\rm SU}(1,1)/\clI$.
To derive this action consider the left action of ${\rm SU}(1,1)$ on $M^+$ (\ref{HC})
$$
M^+\to h\cdot M^+=
=\left(
    \begin{array}{cc}
    A & Az+B \\
    \bar B & \bar Bz+D \\
    \end{array}
  \right)\,.
$$
Due to the Gauss decomposition we need the result of this action up to right
multiplication of the parabolic subgroup $P$ (\ref{x}). The quotient
$(Az+B)(\bar Bz+D)^{-1}$ is $P$-invariant. Thus we get (\ref{mu1}).

In fact, in this action $h$ belongs to the quotient $\bar G=$ SU$(1,1)/(-Id_2)$.
Here $-Id_2$ is the center of SU$(1,1)$

The infinitesimal action of  SU$(1,1)$ (\ref{mu1})  generates the action
 of the Lie algebra (\ref{ls}) on $D$
 \beq{lia}
z\to b+2a z-\bar b z^2\,.
\eq


\subsection{Kuramoto model}

The Lie algebra action (\ref{lia}) generates the the holomorphic flow on the disk $D$
 (\ref{ft0})
\beq{fl1}
\dot{z}=b+2a z-\bar b z^2\,.
\eq
Since the group SU$(1,1)$ acts transitively not only on $D$ but also on its
boundary $S^1$  the flow can be restricted on it. Let $z=e^{\imath\te}$,
Then (\ref{fl1}) becomes
\beq{fl2}
\dot{\te}=-\imath be^{-\imath\te}-2\imath a+\imath \bar be^{\imath \te}\,.
\eq

 Consider an ensemble of $N$ particles
(oscillators) $\te^J$, $J=1,\ldots,N$ arranged as above on the circle and evolving as (\ref{fl2}). Let
\beq{z}
\clZ=\frac{\ka}{N}\sum_{J=1}^Ne^{\imath\te^J}=\frac{\ka}{N}\sum_{J=1}^Nz^J\,.
\eq
be the first moment of the  ensemble and we take $b=\clZ$.
Then we come to the interacting systems of oscillators (\ref{sk0})
\beq{an1}
\dot{z}^I=\clZ+2a z^I-\bar\clZ (z^I)^2\,,~~(I=1,\ldots,N)\,.
\eq
In terms of the phases $\te^I$ it is the system of the Kuramoto equations
\beq{sk}
\dot{\te}^I=
2\imath a+\frac{2\ka}N\sum_{J=1}^N\sin(\te^J-\te^I)\,,~~(I=1,\ldots,N)\,.
\eq
The configuration space for the flow is the $N$-dimensional torus $\times^NS^1$.

Consider the flow on the three-dimensional group $ G$ coming from the Kuramoto flow
(compare with (\ref{fl1}))
\beq{mk0}
\dot{z}^I=\clZ +2az^I -\bar\clZ(z^I)^2\,.
\eq
It can be rewritten as
\beq{mk}
\dot{z}^I=\dot hh^{-1}z^I\,,~~(I=1,\ldots,N)\,, ~~h=
\left(
     \begin{array}{cc}
       A & B \\
       \bar B &  \bar A\\
     \end{array}
   \right)\in G={\rm SU}(1,1)\,.
\eq
Define the flow on $ G$
\beq{gf}
\dot h=
   \left(
     \begin{array}{cc}
       a & \clZ \\
       \bar \clZ &  -a
     \end{array}
   \right)h
   \,,
\eq
or equivalently
 $$
 \left\{
 \begin{array}{c}
   \dot A=a A-\clZ\bar B \\
   \dot B=aB+\clZ\bar A
 \end{array}\right.\,.
$$
It means that if a flow on the three-dimensional group $G$
has the form (\ref{gf}), then (\ref{mk}) is the Kuramoto flow (\ref{mk0})
on the $N$-dimensional configuration space.


\section{Bounded symmetric domains of type I}

In this section we will first define the structure of the domain $D^I_{mn}$
in terms that are applicable to basically all
 classical BSD. Some of them are valid for two exceptional BSD in the Table (see Appendix).
The generalisation  is based on the interrelations between the bounded symmetric domains and
 Hermitian symmetric spaces \cite{He,Wo}.
Then we define the Kuramoto hierarchy of type I.

We use the following notation. Let $A$ be a hermitian matrix. The inequalities
$A>0$, ($A\geq 0$) mean that the matrix $A$ is positive definite (nonnegative definite)
which implies the positivity (nonnegativity) of its leading principal minors.
Equivalently, the eigenvalues of $A$ are positive ((nonnegative).
Let $Z_{mn}$ be the set of matrices with $m$ rows and $n$ columns.
 The domain of the first type is defined as
\beq{ft}
D^I_{mn}=\{\bfz\in Z_{mn}\,|Id_m-\bfz  \bfz^\dag>0\}\,,~~\dim\,D^I_{mn}=2mn\,.
\eq
$$
  Z_{mn}=\{\bfz=(z^\al_j)\}\,,~~\al=1,\ldots,m\,,~j=1,\ldots,n\,.
  $$
We assume that $m\geq n$.



\subsubsection{Hermitian symmetric spaces $X^I$}

Let $G_{mn}=$SU$(m.n)$  be the Lie group from the Table in the Appendix.
For the classical series we consider here $G$ can be represented as a block $2\times 2$ matrices
\beq{G}
G=\left\{g=\left(
             \begin{array}{cc}
               A& B \\
               C & D \\
             \end{array}
           \right)\,~~
\det\,(g)=1\right\}
\eq
$$
A=(A^\al_\be)\,,~B=(B^\al_j)\,,~C=(C_\al^j)\,,~D=(D_j^k)\,.
~\al,\be=1,\ldots m\,,\,j,k=1\ldots,n
$$
These matrices satisfy some additional conditions.
In particular, the group $G_{mn}=$SU$(m.n)$ is fixed by the condition (\ref{sum})
\beq{su}
-AA^\dag +BB^\dag=-I_m\,,~~C=DB^\dag(A^\dag)^{-1}\,,~~DD^\dag-CC^\dag=I_n\,.
\eq

The maximal compact subgroup of $G_{mn}$ is $K_{mn}=$S(U$(m)\times$U$(n))$ (compare with (\ref{k0}))
\beq{k1}
K_{mn}=\left\{
\left( \begin{array}{cc}
               A& 0 \\
               0 & D \\
             \end{array}
           \right)\,~~A\in{\rm U}(m)\,,~D\in{\rm U}(n)
\right\}
\eq
 The quotient space $X^I=G_{mn}/K_{mn}$ is 
noncompact Hermitian symmetric space. Its dual space is the compact Hermitian symmetric space
$X^I_c=U/K_{mn}$
\beq{XI}
X^I=G_{mn}/K_{mn}\,,~~X^I_c= U/K_{mn}\,,~~G_{mn}={\rm SU}(m.n)\,,~~ U={\rm SU}(m+n)\,.
\eq
 The compact space  is the set of $m$-dimensional planes in the complex space
$\mC^{m+n}$ - the Grassmannian Gr$(n,m+n)$ and $\dim\,X^I=\dim\, X_c$. The dual spaces have the same stationary
subgroup $K_{mn}$. On the level of the corresponding Lie algebra the duality is described by the
formulas (\ref{cd}), (\ref{cd1}).

Let $G^\mC=$SL$(m+n,\mC)$. It is the complexification of $G_{mn}$ as well as U.
Represent it in the block form (\ref{G}).
The subgroup $G_{mn}=$SU$(m,n)$ of $G^\mC$ is the fixed set point of the involute automorphism $\si$ ($\si^2=Id$) of $G^\mC$
$$
h\to \si(h)= (sh^\dag s)^{-1}\,,~~s=\di(\underbrace{-1,\ldots,-1}_m,\underbrace{1,\ldots,1}_n)\,.
$$
Note that (\ref{su}) follows from the identity $g=\si(g)$.

The subgroup $U$ is also the fixed set point of another involute automorphism $\te$ of $G^\mC$
$\te(h)=(h^\dag )^{-1}$.

Let $J=(\de_{m+n+1-k,k})$ be the anti-diagonal matrix of order $m+n$. The involution
$h\to Jh^TJ$ transforms the group SU$(m,n)$ into the group SU$(n,m)$. This fact justifies  the
choice $m\geq n$ made above.

The  automorphisms $\te$ and $\si$ commute and $K_{mn}$ is the fixed point set
\beq{K}
K_{mn}=\{h\in G_{mn}\,|\,\te(h)=h\}\sim\{h\in U\,|\,\si(h)=h\}\,,~~K_{mn}={\rm S(U}(m)\times{\rm U}(n))\,.
\eq

The Lie algebra Lie$(G^\mC)=\gg^\mC$ has the block  form
\beq{la1}
\gg=\left\{x=\left(
             \begin{array}{cc}
               a & b \\
               c & d \\
             \end{array}
           \right)\,~~\tr\,a+\tr\,d=0
\right\}\,.
\eq
For $\gg=$Lie(SU$(m,n))$
\beq{pr1}
a^\dag=-a\,,~~ d^\dag=-d\,,~~c=b^\dag\,,
\eq
while for $\gg^c=$Lie$(U)$=SU$(m+n))$ we have $c=-b^\dag$.
Let us denote the action of the involute automorphisms of the Lie algebras  by the same letters.
The action of $\si$ on $\gg$ leads to the Cartan decomposition
\beq{cd}
\gg=\gk\oplus\gm\,,~~\te(\gk)=\gk\,,~\te(\gm)=-\gm\,,
\eq
where $\gk=$Lie$(K_{mn})$, and $\gm$ is the tangent space to $X^I=G_{mn}/K_{mn} $ at the point $K_{mn}$.
Similarly, for $\gu=$Lie(U)
\beq{cd1}
\gu=\gk\oplus\imath\gm\,,~~\si(\gk)=\gk\,,~\si(\gm)=-\gm\,,
\eq
where $\imath\gm$ is the tangent space to the Grassmannian.
In terms of (\ref{la1})
\beq{gk}
\gk=\left(
             \begin{array}{cc}
               a & 0\\
               0 & d \\
             \end{array}
           \right)\,,~~a=-a^\dag\,,~d=-d^\dag\,,~\tr(a+d)=0\,,~~
\gm=\left(
             \begin{array}{cc}
               0 &  b\\
               b^\dag & 0 \\
             \end{array}
           \right)\,.
\eq


\subsubsection{Complexification}

Define the subspace $\gm^+\subset\gg^\mC$
\beq{HC1}
\gm^+=
\left\{\left(
             \begin{array}{cc}
               0 & Z_{mn} \\
               0 & 0 \\
             \end{array}
           \right)
\right\}\,,
\eq
where $Z_{mn}$ is the set of rectangular
matrices (\ref{ft}).

 It can be checked that $\gm^+$ forms a commutative subalgebra of $\gg^\mC$.
 Therefore, the corresponding abelian nilpotent subgroup $M^+$ assumes the form
 $$
 M^+=\exp(\gm^+)=Id_{m+n}+\left(
             \begin{array}{cc}
               0 & Z_{mn} \\
               0 & 0 \\
             \end{array}
           \right)\,.
 $$
 Similarly defined the abelian subgroup
 $$
 M^-=\exp(\gm^-)=Id_{m+n}+\left(
             \begin{array}{cc}
               0 & 0 \\
               Z_{nm} & 0 \\
             \end{array}
           \right)\,.
 $$
Let $K_{mn}^\mC=$S(GL($m,\mC)\times$GL($n,\mC))$ be the complexification of the compact subgroup $K_{mn}$
and $P$ is the parabolic subgroup of $G^\mC$ of the form
$$
P=
\left(
             \begin{array}{cc}
               A &  0\\
               C & D \\
             \end{array}
           \right)\,.
$$
Define the product of three  subgroups (the generalised\emph{ Gauss decomposition})
$$
{\bf X}=M^+K_{mn}^\mC M^-=M^+P\,.
$$
 The product $\bf X$ is  dense in SL($m+n,\mC)$.

 We also need another decomposition the group $G^\mC=$SL$(m+n,\mC)$.  This is the generalised \emph{Iwaswa decomposition}
 \beq{id}
 G^\mC=UP\,,~~({\rm SL}(m+n,\mC)={\rm SU}(m+n)P)\,.
 \eq
 We also have
\beq{id1}
U\cap P=K_{mn}\,,~~({\rm SU}(m+n)\cap P={\rm S(U}(m)\times{\rm U}(n)))\,.
 \eq
 From the last two relations it follows that the Grassmannian Gr$(n,m+n)=X^I_c$ (\ref{XI})
  has an isomorphic description as the quotient $X^I_c=G^\mC/P$. At the same time,
the non-compact space $X^I$ is only included  into  the compact space $G^\mC/P=X^I_c$
$\,(X^I\subset X^I_c)$. It is\emph{ the Borel embedding }generalizing (\ref{be0}).

Let us schematically define \emph{the Harish-Chandra map} for $G_{mn}=$SU$(m,n)$.
Consider the set of rectangular matrices $\gm^+$ (\ref{HC1}).
We identify it with the domain $D^I_{mn}$ imposing constraints
on the matrix $Z_{mn}$ (\ref{ft}).
There is a map $\xi$ of $\gm^+$
to the compact symmetric space $X^I_c=$Gr$(n,m+n)$.
 \beq{HC2}
 \xi\,:\,\gm^+\to X^I_c=G^\mC/P\,, ~~\xi(Z_{mn})=\exp(\gm^+)P\,.
 \eq
It is a diffeomorphism of $D^I_{mn}$  onto dense open subset $X^I$ of $X^I_c$.

The action of group $G_{mn}$ (\ref{G}) on the space $X^I$ corresponds to
 the M\"{o}bius action of $G_{mn}$ (\ref{G})  on the domain $D^I_{mn}$
\beq{mu}
\bfz\to h \bfz=(A\bfz+B)(C\bfz+D)^{-1}\,.
\eq
This form of the $G_{mn}$ action on $D^I_{mn}$ is derived from  its action on $X^I$
in a similar way as was described for the domain $D^I_{11}$ in section 2.
As in that example the center $\clI$ of $G_{mn}$ does not act on $D^I_{mn}$. The center
 is the group of diagonal matrices of the form\\
 $\clI=\{\di(\exp(2\pi\imath j/(m+n))\,,~(j=1,\ldots,m+n)\}$ and we work here with the group
 $G_{mn}/\clI$.

The corresponding action of the Lie algebra $\gg$ (see (\ref{pr1})) takes the form
\beq{la}
\bfz\to J^{\bfz}= b+a\bfz-\bfz d-\bfz b^\dag \bfz\,.
\eq


\subsection{Boundary structure}

The boundary $\p D^I_{mn}$ of the domain $D^I_{mn}$ (\ref{ft}) is the  set defined as
\beq{b}
\p D^I_{mn}=\{\bfz\in  Z_{mn}\,|\,\det(Id_m-\bfz \bfz^\dag)=0\,,~Id_m-\bfz \bfz^\dag\geq 0\}\,,~~\dim(\p D^I_{mn})=2mn-1\,.
\eq


Consider components of the boundary $\p D^I_{mn}$. The exact definition of the components is given in \cite{PS,Wo}.
They have the following description.
 Every component  is the $G_{mn}$-orbits of the  the matrices $\clF^{mn}_{t}$,
 where $1\leq t\leq n$
\beq{ao}
\clF^{mn}_{t}=\left\{
\left(
  \begin{array}{cc}
    Id_t &  \\
     & Z_{m-t,n-t} \\
  \end{array}
\right)\,|\, Id_{m-t}-\bfz  \bfz^\dag>0~for~\bfz\in Z_{m-t,n-t}\right\}\,.
\eq
This form of $\clF^{mn}_{t}$ implies  that the matrix $Id_m-\bfz \bfz^\dag $
(\ref{ft})) has $m-t$ positive eigenvalues and $t$ zero eigenvalues.
It follows from  the Witt theorem \cite{A} that the action of $G_{mn}=$SU$(m,n)$ on
the boundary $\p D^I_{mn}$ preserves  the components.

The subgroup $K_{mn}$  (\ref{mu}) acts  on the boundary $D^I_{mn}=\{Z_{mn}\}$ as
 $\bfz\to A\bfz D^{-1}$, $A\in$U$(m)$, $D\in$U$(n)$.
 By this action $\bfz$ can be "diagonalised"
$$
A\bfz D^{-1}=
\left(
  \begin{array}{cc}
    Id_t & 0 \\
  0   & X_{m-t,n-t} \\
  \end{array}
\right)\,,~~
X_{m-t,n-t}=\left(
    \begin{array}{cccc}
      z_1 & 0 & \cdots & 0 \\
      0 & z_2 & \cdots & 0   \\
      0 &  \cdots &  0 & z_{n-t}  \\
     0 &  \cdots &  0 & 0  \\
      0 &  \cdots &  0 & 0  \\
    \end{array}
  \right)\,.
$$
Thus, $A\bfz D^{-1}\in\clF^{mn}_t$.
 By the definition (\ref{ao}) there is  the projection
 \beq{pr}
 \pi_t\,:\,\clF^{mn}_{t}\to D^I_{m-t,n-t}\,.
 \eq
 Therefore, the components are equivalent to the type I domains $ D^I_{m-t,n-t}$ (\ref{ft}).
 The latter,
 due to the Harish-Chandra maps, are pre-images of the non-compact Hermitian symmetric spaces
 $X^I(t)=$SU$(m-t,n-t)/$S(U$(m-t)\times$U$(n-t))$.
For $t<n$ the components corresponding to the orbits of $\clF^{mn}_{t}$
 are open. The component $\clF^{mn}_n$ is a single closed orbit. It is the \emph{Bergmann-Shilov boundary} $BS(D^I_{mn})$
of the domain $D^I_{mn}$.
This component can be described as the $K_{mn}$-orbit of the matrix $\bfz_0$
\beq{z0}
BS(D^I_{mn})=\left\{ u_1\bfz_0 u_2^{-1}\,|\,
\bfz_0=\left(
  \begin{array}{c}
    Id_n   \\
  0 \\
  \end{array}
\right)\,,~~u_1\in{\rm U}(m)\,,~u_2\in{\rm U}(n)\right\}\,.
\eq
It follows from this definition that the BS boundary is the quotient
\beq{sm}
 BS(D^I_{mn})={\rm U}(m)/{\rm U}(m-n)\,,~~\dim\,BS(D^I_{mn})=n(2m-n)\,.
 \eq
 This quotient is \emph{ the complex Stiefel manifold} $St^\mC_{mn}$.
 The Stiefel manifold is the set of all orthonormal n-frames in the $m$-dimensional space $\mC^m$.
It is a compact homogeneous space,
which, as we shall see, defines the configuration space
of the Kuramoto model.
 Equivalently, the component $BS(D^I_{mn})$ is defined by the condition
\beq{bs1}
St^\mC_{mn}=\{\bfz\in D^I_{mn}\,|\, Id_n-\bfz^\dag\bfz=0\}\,.
\eq
This condition is not the same as $Id_m-\bfz\bfz^\dag=0$ unless $m=n$.
It follows from (\ref{sm}) that  for $m=n$ the Stiefel manifold
is the space of unitary matrices $St^\mC_{mm}\sim\,$U$(m)$ and for $n=1$ it is the odd-dimensional sphere
$St^\mC_{mm}\sim S^{2m-1}$.
 Summarizing,
 the topological boundary is the union of the components
 \beq{ub}
 \p D^I_{mn}=\bigcup_{t=1}^{n}G_{mn}(\clF^{mn}_t)\,.
 \eq
 In the following to denote the components we will use their $G$-equivalence classes $\clF^{mn}_t$.

 Our main interest are the BS boundaries. They have the following description.
   By means of (\ref{pr}) put in the correspondence to the component $\clF_t^{mn}$ the domain $D^I_{m-t,n-t}$.
    Consider the boundaries of $D^I_{m-t,n-t}$. They are the $G_{m-t,n-t}$ orbits of the matrices
    $\clF_{t_1}^{m-t,n-t}$\\
     $(t_1=1,\ldots,n-t)$ and $(t=1,\ldots,n-1)$.
    All but one of the components are open. The closed component   $\clF_{n-t}^{m-t,n-t}$ corresponds to $t_1=n-t$.
    It is the BS boundary of the domain $BS(D^I_{m-t,n-t})=St^\mC_{m-t,n-t}$. This boundary
    is the $K_{m-t,n-t}$ orbit of the matrix $\bfz_0\in Z_{m-t,n-t}$ (compare with (\ref{z0}))
    \beq{z1}
BS(D^I_{m-t,n-t})=\left\{ u_1\bfz_0 u_2^{-1}\,|\,
\bfz_0=\left(
  \begin{array}{c}
    Id_{n-t}   \\
  0 \\
  \end{array}
\right)\,,~~u_1\in{\rm U}(m-t)\,,~u_2\in{\rm U}(n-t)\right\}\,.
\eq
  Thus,  the domain $D^I_{mn}$ has the chain of closed BS boundaries
    \beq{bc}
BS(D^I_{mn})\,,\,BS(D^I_{m-1,n-1})\,,\ldots,\,BS(D^I_{m-n+1,1})\,.
\eq
The boundaries $\p \clF^{mn}_t$ of $\clF^{mn}_t$ are the boundaries of $D^I_{mn}\,$:
$\,\p D^I_{mn}\supset\p \clF^{mn}_t\,,~~t=1,\ldots,n$.


\subsection{ Kuramoto families}

Define a flow on the domain $D^I_{mn}$ (\ref{ft}) corresponding to the Lie algebra action  (\ref{la})
$\dot \bfz=J^z\bfz$
\beq{f21}
\dot \bfz=b+a\bfz-\bfz d-\bfz b^\dag \bfz\,.
\eq
Here $(a,b,d)\in\gg=$Lie(SU$(m,n))$ (\ref{la1}).
The equation (\ref{f21}) has the form of the matrix Riccati equation.
%

Consider $N$  oscillators  $\bfz^K\,,~~K=1,\ldots N$ on the bounded domain  $D^I_{mn}$.
In other words we pass to the configuration space
$\clX=\times_{K=1}^N(D_{mn}^I)^K$ and to the $K$ flows
\beq{f2}
\dot \bfz^K=b+a\bfz^K-\bfz^K d-\bfz^K b^\dag \bfz^K\,, ~~K=1,\ldots N\,.
\eq
It is a system of non-interacting oscillators. In order for the oscillators to interact with each other
 we assume that the matrices $a$, $b$ and $d$  depend on some mean-field quantities.
In particular, consider the moment
\beq{an}
\clZ=\frac{\ka}{N}\sum_{J=1}^N\bfz^J\,.
\eq
and take $b=\clZ$ while $a$ and $d$ remain $\bfz^J$-independent.
The system (\ref{f2}) becomes
\beq{f3}
\dot \bfz^K=a\bfz^K-\bfz^Kd+\frac{\ka}{N}\sum_{J=1}^N(\bfz^J-\bfz^K (\bfz^\dag)^J \bfz^K)\,.
\eq
\beq{f4}
\frac{d}{dt}( z^\al_j)^K=a^\al_\be(z^\be_j)^K-(z^\al_k)^Kd^k_j+
\frac{\ka}{N}\sum_{J=1}^N((z^\al_j)^J-(z^\al_k)^K(z^k_\be)^J(z^\be_j)^K)
\eq
with initial data $(z^\al_j)^K(t)_{t=0}=(z^\al_j)^K(p)$.
The Einstein summation rule is assumed in (\ref{f4}).
We call the system (\ref{f4})\emph{ the type I Kuramoto model $KM_{mn}(I)$},
or taking into account the dual symmetric space the Grassmannian model Gr$_{n,m+n}$.
For $m=n=1$ put $z^K=e^{\imath\te^K}$. Then the system (\ref{f4}) coincides with the Kuramoto system (\ref{sk0}).
This system is the $\mC P^1$ model.

Due to the Witt theorem \cite{A} the group $G=$SU$(m,n)$ as well as its subgroup SU$(m-t,n-t)$ acts
  on the components
$\clF^{mn}_{t}\sim D_{m-t,n-t}$,(\ref{ao}) of the
boundary.
One can restrict the flow (\ref{f4}) on them. We
 assume that the coefficients $a,b,d$ in (\ref{f2}) belongs to the
subalgebra Lie$(G_{m-t,n-t})$.
 These components are in general open. But the Kuramoto type flows should be defined
on  compact spaces. For this reason we should restrict the flows on the defined above the BS boundaries or the Stiefel manifolds  (\ref{bc}).
Thus we come to  the family of $n$ KM(I) models related to the domain $D^I_{mn}$
\beq{hk1}
KM_{mn}(I)\,,\, KM_{m-1,n-1}(I)\,,\ldots,KM_{m-n+1,1}(I)\,.
\eq
Their configuration spaces are the product of the Stiefel manifolds
$\times_{J=1}^{N} (St^\mC_{m-t,n-t})$.
The equations of motion have the form (\ref{f3}), where $\bfz^K$ are local coordinates
on the Stiefel manifold $St^\mC_{m-t,n-t}=$U$(m-t)/$U$(m-n)$ $(t<n)$.

We can pass to some fixed element $ KM_{m-t,n-t}$ of the family.  For this purpose assume that the initial configuration belongs to some component of the boundary
$(z^\al_j)^I(p)\in D_{m-t,n-t}$.
It means that for this configuration the group of symmetries $G_{m,n}$
is broken up to the subgroup $G_{m-t,n-t}$.
In this case the flows remain localized at the component
 $D^I_{m-t,n-t}$. The flows on its BS boundary are just $ KM_{m-t,n-t}(I)$ i.e on the
 Stiefel varieties $St^\mC_{m-t,n-t}$, $\,\dim\,(St^\mC_{m-t,n-t})=(n-t)(2m-n-t)$.
The symmetry can be broken down step-by-step
$$
G_{mn}\to G_{m-t_1,n-t_1}\to\ldots\to G_{m-t_k,n--t_k}\,.
$$
If $m=n$ the Stiefel manifold is the unitary group $St^\mC_{mn}=$U$(m)$,
the Kuramoto family is the Lohe unitary family. The end of the family is the standard
Kuramoto model $KM_{1,1}(I)$. If $n<m$ the family  ends with
the $\mC P^{n}$ model or $KM_{m-n,1}(I)$, or the spherical $S^{2n-1}$ model.

Consider the Kuramoto flows $KM_{m-t,n-t}(I)$ (\ref{f3}).
 It can be rewritten in term of the group element  $h\in G_{m-t,n-t}$
\beq{mki}
\dot{z}^K=\dot hh^{-1}z^K\,,~~(K=1,\ldots,N)\,,~~\bfz^K\in St^\mC_{m-t,n-t}\,,
\eq
if  the flow on  $G_{m-t,n-t}$   has the form
\beq{gf1}
\dot h=xh\,, ~~
x=\left(
             \begin{array}{cc}
               a & \clZ \\
               \clZ^\dag & d \\
             \end{array}
           \right)\,~~(\tr\,a+\tr\,d=0)\,.
\eq
In this way we come to the reduction from the $N$ flows on the Stiefel manifold
$St^\mC_{m-t,n-t}$ to the flow on the group $G_{m-t,n-t}$
$$
\times_{J=1}^{N} (St^\mC_{m-t,n-t})\to G_{m-t,n-t}\,.
$$
Note that
$\dim\,(\times_{J=1}^{N} (St^\mC_{m-t,n-t}))=N(n-t)(2m-n-t)$ and\\
$\dim\,(G_{m-t,n-t})=(m+n)^2-4t(m+n)+4t^2-1$.

\subsection{Examples}
\subsubsection{$\mC P^m$ model}

This example
 is an evident generalization of the SU$(1,1)$ case.
We consider it very briefly since it has been analysed in \cite{CEM,Cr,LMS,Ta}.

Using (\ref{ft}) we define the bounded domain for $n=1$
$$
D^I_{m1}=\{\bfz^\dag=(\bar z_1,\ldots,\bar z_m)\in\mC^m\,|\,Id_m-\bfz  \bfz^\dag>0\}\,,
$$
This domain is the unit open ball in $\mC^m: \,\sum_{k=1}^n|z_k|^2<1$.
The Harish-Chandra embedding of this ball in $\mC P^m$ means that its image is
 the noncompact symmetric space\\
 $X^I=$SU$(m,1)/$S(U$(m)\times$U$(1))$ dual to
  $X^I_c=\mC P^m=$SU$(m+1)/$S(U$(m)\times$U$(1))$.

The only component of the boundary $\p D^I_{m1}$ is the sphere $S^{2m-1}=\{\bfz\,|\,\sum_{\al=1}^m\,|\,z^\al|^2=1\}$. 
 It is the BS boundary and it is nothing but the Stiefel manifold $St^\mC_{m,1}$.

 The Kuramoto system (\ref{f2}) on $S^{2m-1}$ takes the following form.
 Let $\bfz^K$, $(K=1,\ldots,N)$ be the set of $N$ points on the sphere $S^{2m-1}$.
 In this case the matrix $a$ is a $m\times m$ matrix, $b$ is a column $m$-vector
$ \frac{\ka}{N}\sum_{J=1}^N\bfz^J$ and  $d$ is a scalar.
 Then the Kuramoto flows on $S^{2m-1}$ are
 $$
 \frac{d}{dt}\bfz^K=(a-d)\bfz^K+
\frac{\ka}{N}\sum_{J=1}^N(\bfz^J-\bfz^K\bfz^J\bfz^K)\,.
 $$
 As in the general case they are equivalent to the flow on the group SU$(m,1)$ (\ref{gf1}).

\subsubsection{Gr$(2,4)$ model}
We consider the domain (\ref{ft}) for $m=n=2$.
Let $\bfz$ be  the matrix
$$
\bfz=
\left(
  \begin{array}{cc}
    z_{11} & z_{12} \\
    z_{21} & z_{22} \\
  \end{array}
\right)\in Z_{22}\,.
$$
The bounded domain corresponding to this case is defined as
$$
D^I_{2,2}=\{\bfz\in Z_{22}\,|\,Id_2-\bfz \bfz^\dag>0\}\,.
$$
 The matrix elements satisfy two inequalities
\begin{subequations}\label{TS}
  \begin{align}
 & |z_{11}|^2+|z_{12}|^2<1\,,\\
 &\sum_{jk=1}^2|z_{jk}|^2+
 |z_{12}|^2|z_{21}|^2+|z_{11}|^2|z_{22}|^2 -(\bar z_{12}\bar z_{21}z_{11}z_{22}+c.c)<1\,.
 \end{align}
  \end{subequations}
  The Harish-Chandra maps is $\xi\,:\,D^I_{2,2}\to Gr(2,4)={\rm SU}(4)/S({\rm U}(2)\times{\rm U}(2))$.

The boundary  $\p D^I_{2,2}$ is  defined by the conditions
$$
\p (D^I_{2,2})=(\{\bfz\,|\det\,(Id_2-\bfz \bfz^\dag)=0\,,~Id_2-\bfz \bfz^\dag\geq 0
)=0\,, ~~\dim\,\p (D_{22}^I)=7\,.
$$
It means that the inequality (\ref{TS}b) is replaced on equality.

The BS boundary $BS(D_{22}^I)$ of $D^I_{2,2}$  is defined by the equation $Id_2-\bfz^\dag \bfz=0$ (\ref{bs1}).
As in the general case this condition implies that $\bfz$ is a unitary matrix U$(2)$.
Thus, the BS boundary is the group U$(2)$ and $\dim\,BS(D_{22}^I)=4$.

There is the open component of the boundary containing the matrix $\clF_1^{11}\sim D^I_{1,1}$ (\ref{ao})
$$
\left(
  \begin{array}{cc}
    1 & 0 \\
    0 & z_{22} \\
  \end{array}
\right)\,,~~|z_{22}|<1\,.
$$
and  $D^I_{1,1}$
is the  unit disc in $\mC P^1$ (\ref{ld}).
 The BS boundary $BS( D^I_{1,1})$ is the
unit circle $S^1$.
So in this case the KM family consists of two Kuramoto models.
One is the standard KM on $S^1$ and the other is the U$(2)$ Lohe model.

More generally, for the Gr$(2,m)$ model $G_{m2}={\rm SU}(m,2)$, the BSD $D^I_{m,2}$ is defined by
the inequality $Id_m-\bfz\bfz^\dag>0$ for $\bfz\in Z_{m,2}$.
It has two BS boundary components. The first one is the Stiefel manifold $St^\mC_{m,2}={\rm U}(m)/{\rm U}(2)$.
The second is the sphere $St^\mC_{m-1,1}=S^{2m-3}$. So this generalisation leads to the two
terms Kuramoto family related to  the Stiefel manifold $St^\mC_{m,2}$ and the spheres $S^{2m-3}$.


\section{Classical domains of type  II}

Consider the domain of the first type (\ref{ft}).
Assume that  $m=n$ and the matrices $\bfz$ are antisymmetric.
Let $Z_{n}$ be the set of complex matrices of order $n$. The domains of type II are defined as
\beq{ii}
D_n^{II}=\{\bfz\in Z_{n}\,|Id_n-\bfz^\dag  \bfz>0\,,~\bfz=-\bfz^T\}\,,~~\dim_\mC\,D^{II}_{n}=\frac{n(n-1)}2\,.
\eq
For $n=1$ this domain is just a point and we assume that $n>1$.

\subsection{Symmetric space $X^{II}$}

Consider the symmetric space $X^{II}$ corresponding to the domain $D_n^{II}$. This is
 non-compact Hermitian symmetric space $X^{II}=G/K$
  with $G=$SO$^*(2n)$ and $K=$U$(n)$.
The group SO$^*(2n)$ is a real form of the complex group
SO$(2n,\mC)$. We define the latter  as the matrices from SL$(2n,\mC)$ preserving the
symmetric form $s_1$
$$
gs_1g^T=s_1\,,~~s_1=\left(
  \begin{array}{cc}
    0 & Id_n \\
    Id_n & 0 \\
  \end{array}
\right)\,.
$$
It means that the matrices from SO$(2n,\mC)$ have the form
\beq{so2}
g=\left(
  \begin{array}{cc}
    A & B \\
    C & D \\
  \end{array}
\right)\,,~~A^TC=-C^TA\,,~D^TB=-B^TD\,,~A^TD+C^TB=Id_n\,.
\eq

The real subgroup SO$^*(2n)$ obeys the additional constraints
\beq{ac}
gs_2g^\dag=s_2\,,~~s_2=\left(
  \begin{array}{cc}
 -  Id_n  & 0\\
   0 & Id_n  \\
  \end{array}
\right)\,,
\eq
which means that
$$
AA^\dag-BB^\dag=Id_n\,,~~DB^\dag-CA^\dag=0\,,~~DD^\dag-CC^\dag=Id_n\,.
$$

So SO$^*(2n)=$SO$(2n,\mC)\cap$U$(n,n)$.
In this way the group SO$^*(2n)$ is a fixed point set in the group SO$(2n,\mC)$ under the action of the involutive automorphism $\si$
$$
h\to \si(h)= (s_2h^\dag s_2)^{-1}\,.
$$

The maximal subgroup of SO$^*(2n)$ is $K=$U$(n)$. It follows from (\ref{so2}) that
\beq{k2}
K=\left\{\left(
  \begin{array}{cc}
    A & 0 \\
   0  & \bar A\\
  \end{array}
\right)\,,~A\in{\rm U}(n)
\right\}\,.
\eq
Define the noncompact symmetric space $X^{II}$ as the quotient $X^{II}=$ SO$^*(2n)/$U$(n)$.
The dual compact space is $X_c^{II}=$ SO$(2n)/$U$(n)$. The group SO$(2n)$ is the compact real form of
 SO$(2n,\mC)$. It is invariant with respect to the involutive automorphism $\te$:
$\te(h)=\bar h$. Thus, for the type II domains we have
\beq{ss2}
X^{II}={\rm SO}^*(2n)/{\rm U}(n)\,,~~X_c^{II}={\rm SO}(2n)/{\rm U}(n)\,.
\eq

Define the maximal parabolic subgroup of SO$(2n,\mC)$
\beq{p2}
P=\left\{\left(
  \begin{array}{cc}
    A & 0 \\
    C & \bar A \\
  \end{array}
\right)
\right\}\,,~A^TC=-C^TA\,,~~(\,P\cap K= K)
\,.
\eq

Let
$$
\gm^+=\left\{\left(\begin{array}{cc}
    0 & Z_n \\
    0 & 0 \\
  \end{array}
\right)\,,~~Z_n^T=-Z_n\right\}
$$
and we assume that $ Z_n$ satisfies (\ref{ii}).
 The Harish-Chandra map is  $\xi\,:\,\gm^+\to X_c^{II}$ $\,(\xi(Z_n)=\exp(\gm^+)P)$. Its image is the non-compact space $X^{II}$.
 The group SO$^*(2n)$ acts on the domain  $D^{II}$ by the  M\"{o}bius transformation (\ref{mu}).
 Again we pass to the quotient SO$^*(2n)/\clI$, where $\clI$ is the $\mZ_2$
  center $\clI=\{\pm Id_{2n}\}$ of SO$^*(2n)$.

The Lie algebra Lie(SO$^*(2n))=$so$^*(2n)$ is represented by the matrices
\beq{la2}
\gg=\left\{x=\left(
             \begin{array}{cc}
               a & b \\
               b^\dag & \bar a \\
             \end{array}
           \right)\,,~~a=-a^\dag\,,~b=-b^T
         \right\}\,.
\eq
It acts on the domain $D^{II}_{n}$ (\ref{ii}) as the vector field $\bfz\to J^z\bfz$,
\beq{vf5}
J^z=b+a\bfz-\bfz\bar a-\bfz b^\dag\bfz\,.
\eq

%
\subsection{The boundary structure}

The boundary $\p D_n^{II}$ of the domain $D_n^{II}$ (\ref{ft}) is defined as
\beq{b2}
\p D_n^{II}=\{\bfz\in  Z_n\,|\,\det(Id_n-\bfz \bfz^\dag)=0\,,~Id_n-\bfz \bfz^\dag\geq 0\}\,.
\eq

The group $G=$SO$^*(2n)$ acts on $\p D_n^{II}$ by the M\"{o}bius transformation (\ref{mu}).
The boundary contains $l$ components.
where $l=(n-1)/2$ for odd $n$ and $l=n/2$ for even $n$.
Any  component  contains the matrix
\beq{c2}
\clF_t=\left\{
\left(
  \begin{array}{cc}
    J_t & 0 \\
  0   & Z_{m-2t} \\
  \end{array}
\right)\,,~~J_t=
\left(
  \begin{array}{cc}
    0 & Id_t \\
  -Id_t  & 0 \\
  \end{array}
\right)
\right\}\,.
\eq
It follows from (\ref{ii}) that this component can be projected as for the type I domains (\ref{pr})
 to the bounded domain $D_{n-2t}^{II}$.
The action of  the subgroup $K$  (\ref{k2})
$A\clF_tA^T$
 is transitive on the $\clF_t$-component.

Assume that $n=2l$ is even.
As above, the component $G(\clF_l)$ is closed. It is the BS boundary $BS(D^{II}_{2l})$.
To describe it consider the $K$-orbit of the element $\clF_l=J_n$
 $$
 BS(D^{II}_n)=\{u\in {\rm U}(n)\,|\,uJ_nu^T\}\,.
 $$
The stability subgroup  of this action is Sp$(l)$.
Thus,
$$
BS(D^{II}_{2l})={\rm U}(2l)/{\rm Sp}(l)\,,~~
\dim\,BS(D^{II}_{2l})=2l^2-l=\frac{n(n-1)}2\,.
$$
Since $J^T_n=-J_n$ the BS boundary is the space of unitary antisymmetric matrices
\beq{bs9}
BS(D^{II}_{2l})=\{u\in{\rm U}(2l)\,|\,u^T=-u\}\,.
\eq

Let $n=2l+1$. To define the BS boundary consider the $K$-orbit of the matrix
$$
\clF_l=
\left(
  \begin{array}{cc}
    J_l & 0 \\
  0   & 0 \\
  \end{array}
\right)\,.
$$
The stability subgroup  of this action is Sp$(l)$.  Therefore,
$$
 BS(D^{II}_{2l+1})=\{u\in {\rm U}(2l+1)\,|\,u\clF_lu^T\
$$
Since $\clF_l^T=-\clF_l$ the boundary $BS(D^{II}_{2l+1})$  is the space of unitary antisymmetric matrices
as in the even case (\ref{bs9}) and $\dim\,(BS(D^{II}_{2l+1}))=\frac{n(n-1)}2$.

Similar,  the components corresponding to $\clF_l$ for $l<n/2$ or $l<(n-1)/2$ have their BS boundaries.
In this way we have two families of the
BS boundaries
\beq{bs2}
\begin{array}{ccccccc}
           \p D^{II}_{2l}\,:& & BS(D^{II}_{2l})\,,& &BS(D^{II}_{2l-2})\,,  & \cdots & BS(D^{II}_{2})\,, \\
           &&&&\\
           \p D^{II}_{2l+1}\,:&& BS(D^{II}_{2l+1})\,,&  &BS(D^{II}_{2l-1})\,,  & \cdots & BS(D^{II}_{1})\,.
         \end{array}
\eq
Note that $BS(D^{II}_{2})=$U$(2)/$Sp$(1)\sim S^1$.


\subsection{ Kuramoto families}

The flow on the domain $D^{II}_{n}$ (\ref{ii}) corresponding to the Lie algebra action  $J^z$ (\ref{vf5})
\beq{fz5}
\dot \bfz=b+a\bfz-\bfz\bar a -\bfz b^\dag\bfz\,,
\eq
where $\bfz$ is an antisymmetric complex matrix of order $n$ satisfying (\ref{ii}).
As we have already done for the domains of type I, we take $b=\clZ$ (\ref{an}).
 Then from  (\ref{fz5}) we come to a
system similar to (\ref{f3}) with $\bfz\in D^{II}_{n}$
\beq{kii}
\dot \bfz^K=a\bfz^K-\bfz^K\bar a+\frac{\ka}{N}\sum_{J=1}^N(\bfz^J-\bfz^K (\bfz^\dag)^J \bfz^K)\,.
\eq
Recall that the group $G={\rm SO}^*(2n)$ acts on the components of the boundary $\p D^{II}_{n}$.
Thereby we can restrict the flow (\ref{kii}) on the chains of the BS boundaries (\ref{bs2}).
We come to the two families of the type II  Kuramoto models for odd and even $n$
$$
KM_{2l+1}(II)\,,\, KM_{2l-1}(II)\,,\ldots,KM_{1}(II)\,,
$$
$$
KM_{2l}(II)\,,\, KM_{2l-2}(II)\,,\ldots,KM_{2}(II)\,.
$$
Their configuration spaces are $\times_{J=1}^N (BS(D^{II}_{2l+1-t}))$  for $KM_{2l+1-t}(II)$
and $\times_{J=1}^N (BS(D^{II}_{2l-t}))$  for $KM_{2l-t}(II)$.
Since BS boundaries are  unitary antisymmetric matrices the type II Kuramoto families are the
Lohe families invariant under the reduction $u\to -u^T$ (\ref{bs9}).

As for the type I Kuramoto models
there are reductions of these flows to the flows on the group manifolds
$\times_{J=1}^N (BS(D^{II}_{2l+1-2t}))\to{\rm SO}^*_{4l+2-2t}$ and
$\times_{J=1}^N (BS(D^{II}_{2l-2t}))\to{\rm SO}^*_{4l-2t}$.


\subsubsection{Examples}

Consider particular cases for small $n$.\\
{\bf 1.}$KM_{2}(II)$\\
The corresponding symmetric space is $X^{II}=$SO$^*(4)/$U$(2)$.
The group SO$^*(4)$ is isomorphic to the product SU$(2)\times$SU$(1,1)$ \cite{He}.
Since SU$(2)=$U$(2)/$U$(1)$ the symmetric space in this case takes the form
  $X^{II}=$SU$(1,1)/$U$(1)$. This symmetric space corresponds to the standard Kuramoto model
(see Section 1) or $KM_{1,1}(I)$ (\ref{hk1}).\\
{\bf 2.}$KM_{3}(II)$\\
In this case SO$^*(6)\sim$SU$(3,1)$ and we come to the symmetric space\\
$X^{II}\sim X^I=$SU$(3,1)$/S(U$(3)\times$U$(1))$  \cite{He}. The dual compact space is $X^I_c=\mC P^3$
  So $KM_{3}(II)\sim KM_{3,1}(I)$.
It is the spherical $S^5$ model or the $\mC P^3$ model considered above.\\
{\bf 3.}$KM_{4}(II)$\\
In this case the noncompact symmetric space is $X^{II}=$SO$^*(8)/$U$(4)$.
Since\\
 SO$^*(8)/$U$(4)\sim$SO$(6,2)/$SO$(6)\times$SO$(2)$
the Kuramoto model $KM_{4}(II)$ is isomorphic to a type IV model,
which we not consider here.
In our case the chain of the BS boundaries has two terms $BS(D^{II}_{4})\,,~ BS(D^{II}_{2})$ (\ref{bs2})
or ${\rm U}(4)/{\rm Sp}(2)$, ${\rm U}(2)/{\rm Sp}(1)$.
The Kuramoto family\\ $(KM_{4}(II)\,,~ KM_{2}(II))$ is ended with the
standard Kuramoto model.

\section{Classical domains of type  III}
 The domain of the  type III is similar to the domain of the
 type II (\ref{ii}) where the antisymmetric  matrices are replaced
 on symmetric matrices
\beq{iii}
D_n^{III}=\{\bfz\in Z_{nn}\,|Id_n-\bfz^\dag  \bfz>0\,,~\bfz=\bfz^T\}\,,~~\dim_\mC\,D^{III}_{n}=\frac{n(n+1)}2\,.
\eq

\subsection{Symmetric space $X^{III}$}

The non-compact Hermitian symmetric space we need
is the quotient $X^{III}=G/K$,
 where $G$ is a real symplectic group Sp$(n,\mR)$ and $K=$U$(n)$.
The group Sp$(n,\mR)$ is a real form of the complex group
Sp$(n,\mC)$. We define the latter  as the matrices from SL$(2n,\mC)$ preserving the
symmetric form $s_1$
$$
gs_1g^T=s\,,~~s_1=\left(
  \begin{array}{cc}
    0 & Id_n \\
    -Id_n & 0 \\
  \end{array}
\right)\,.
$$
\beq{G3}
{\rm Sp}(n,\mC)=
\left\{g=\left(
  \begin{array}{cc}
    A & B \\
    C & D \\
  \end{array}
\right)\,,~~A^TC=C^TA\,,~D^TB=B^TD\,,~A^TD-C^TB=Id_n\right\}\,.
\eq
Consider the parabolic subgroup $P$ of ${\rm Sp}(n,\mC)$
$$
\left\{\left(
  \begin{array}{cc}
    A & 0 \\
    C & D \\
  \end{array}
\right)\,,~~A^TC=C^TA\,,~A^TD=Id_n\right\}\,.
$$
Its maximal compact subgroup is
\beq{k5}
K=\left\{\left(
  \begin{array}{cc}
    A & 0 \\
    0 & \bar A \\
  \end{array}
\right)\,,~~A\in{\rm U}(n)\right\}\,.
\eq

The maximal compact real form of ${\rm Sp}(n,\mC)$ is
${\rm Sp}(n)={\rm Sp}(n,\mC)\cap{\rm U}(2n)$. To define the domain $D^{III}$
we need the noncompact form
$$
G={\rm Sp}(n,\mR)={\rm Sp}(n,\mC)\cap{\rm U}(n,n)\,.
$$
It follows from (\ref{ac} that
$$
G=\{g\in{\rm Sp}(n,\mC)\,|\,gs_2g^\dag=s_2\}\,.
$$
It means that in addition to (\ref{G3}) $G$ satisfies the constraints
$$
AA^\dag-BB^\dag=Id_n\,,~~DB^\dag-CA^\dag=0\,,~~DD^\dag-CC^\dag=Id_n\,.
$$

The Lie algebra $\gg=$Lie($G)={\rm sp}(n,\mR)$ has the form
\beq{la3}
\gg=\left\{x=\left(
             \begin{array}{cc}
               a & b \\
               b^\dag & \bar a \\
             \end{array}
           \right)\,,~~a=-a^\dag\,,~~b=b^T
         \right\}\,.
\eq

The noncompact Hermitian symmetric space is the quotient
$$
X^{III}=G/K=
{\rm Sp}(n,\mR)/U(n)
$$
while the dual compact space is $X_c^{III}={\rm Sp}(n)/U(n)$.
Let
$$
\gm^+=\left\{\left(\begin{array}{cc}
    0 & Z_n \\
    0 & 0 \\
  \end{array}
\right)\,,~~Z_n^T=Z_n\right\}
$$
and we assume that $ Z_n$ satisfies (\ref{iii}).
The Harish-Chandra map is $\xi\,:\,\gm^+\to X_c^{III}$
and the image of  $D^{III}_{n}$ is $X^{III}$. It is an open dense
subset in $X^{III}_c$.
The Group Sp$(n)$  acts on the domain  $D^{III}$ by the  M\"{o}bius transformation (\ref{mu}).
 We pass to the quotient Sp$(n)/\clI$, where $\clI$ is the $\mZ_2$
  center $\clI=\{\pm Id_{2n}\}$ of Sp$(n)$.
  The action of the Lie algebra on the domain $D^{III}_{n}$ (\ref{iii}) has as above
  the four terms form
\beq{vf2}
\bfz\to J^z=b+a\bfz-\bfz\bar a-\bfz b^\dag\bfz\,,
\eq
where $b$ is a symmetric matrix.

%
\subsection{The boundary structure}

The boundary $\p D_n^{III}$ of the domain $D_n^{III}$ (\ref{ft}) is the set
$$
\p D_n^{III}=\{\bfz\in  Z_n\,,~\bfz=\bfz^T\,|\,\det(Id_n-\bfz \bfz^\dag)=0\,,~Id_n-\bfz \bfz^\dag\geq 0\,,\}\,.
$$
The group $G$ acts transitively  on $\p D_n^{III}$ by the M\"{o}bius transformation (\ref{mu}).
The boundary has $n$ components. They are $G$-orbits of the matrices $\clF_t$, $(t=1,\dots n)$
\beq{c3}
\clF_t=\left\{
\left(
  \begin{array}{cc}
    Id_t & 0 \\
  0   & Z_{n-t} \\
  \end{array}
\right)
\right\}\,.
\eq
It follows from (\ref{iii}) that the component is holomorphically equivalent to the bounded domain $D_{n-t}^{III}$. All orbits for $t\neq n$ are open. The orbit $G(\clF_n)$ is
closed. It is the BS boundary of the domain  $D_n^{III}$.
The subgroup $K$  (\ref{k5}) acts transitively on it.
$$
A(Id_n) A^T\,,~~A\in{\rm U}(n)\,.
$$
The stability subgroup  of this action is the orthogonal subgroup O$(n)$.
Thus, $BS(D^{III}_{n})$ is the space of unitary symmetric matrices
\beq{bs3}
BS(D^{III}_{n})={\rm U}(n)/{\rm O}(n)=\{u\in{\rm U}(n)\,|\,u=u^T\}\,,~~\,.
\eq
and $\dim\,BS(D^{III}_{n})=\frac{n(n+1)}2$.


Because the components $\clF_t$ for $t<n$ are isomorphic to the domains $D^{III}_{n-t}$
 we have the chain of the BS boundaries
\beq{bs4}
           \p D^{III}_{n}\,:\,~ BS(D^{III}_{n})~  BS(D^{III}_{n-1}) \cdots  BS(D^{III}_{1})
\eq


\subsection{ Kuramoto families}

The flow on the domain $D^{III}_{n}$ (\ref{ii}) corresponding to the Lie algebra action  $J^z$ (\ref{vf2})
\beq{fz}
\dot \bfz=b+a\bfz-\bfz\bar a -\bfz b^\dag\bfz\,,
\eq
where $\bfz$ is an symmetric complex matrix of order $n$ satisfying (\ref{ii}).
As we have already done for the domains of type I and II, we take $b=\clZ$ (\ref{an}).
 Then the equation (\ref{fz}) become the
system (\ref{f3}) with $\bfz\in D^{III}_{n}$
\beq{k3}
\dot \bfz^K=a\bfz^K-\bfz^K\bar a+\frac{\ka}{N}\sum_{J=1}^N(\bfz^J-\bfz^K (\bfz^\dag)^J \bfz^K)\,.
\eq
The  configuration space for the $N$ particle system is $\times_{J=1}^N (BS(D^{III}_{n-t}))$.
It has dimension $\frac{Nn(n=1)}2$. On the other hand, as above, the system (\ref{k3}) compatible with the flow on the group ${\rm Sp}(n,\mR)$ (\ref{gf1}). The configuration space for this flow has dimension $n(2n+1)$.

Restricting the flows (\ref{k3}) on the BS boundaries (\ref{bs4}) we come to the  families of the type III  Kuramoto models
$$
KM_{n}(III)\,,\, KM_{n-1}(III)\,,\ldots,KM_{1}(III)\,.
$$
The last member in the family is the standard Kuramoto model (see below).
The configuration spaces are $\times_{J=1}^N (BS(D^{III}_{n-t}))=\times_{J=1}^N {\rm U}(n-t)/{\rm O}(n-t)$.
The BS boundaries are the unitary symmetric matrices (\ref{bs3}) of order $n-t$. Therefore, the type III Kuramoto families are the
Lohe families invariant under the reduction $u\to u^T$ (\ref{bs9}).


\subsubsection{Examples}

{\bf 1.}$KM_{1}(III)$\\
It is the standard Kuramoto model, because $G={\rm Sp}(1,\mR)\sim{\rm SU}(1,1)$,
 $D^{III}_1$ is the unit disc and the BS boundary is the circle $S^1$.\\
{\bf 2.}$KM_{2}(III)$\\
The BS boundary is ${\rm U}(2)/{\rm O}(2)$. It is isomorphic to the sphere $S^3\sim\rm SU(2)$.
The family includes also the $S^1$ model $KM_{1}(III)$.
Since $G={\rm Sp}(2,\mR)/\mZ_2\sim{\rm SO}(3,2)$
the domain $D^{III}_{2}$ is isomorphic to a domain of type IV related to the space
$X={\rm SO}(3,2)/{\rm SO}(3)\times{\rm SO}(2)$.

\section{Concluding remarks}

There are few aspects that have not discussed here.

It would be interesting to understand whether such properties
 as synchronisation phenomena hold in general for the type I, II and III Kuramoto models.
  It seems that the simplest case to test this is  the type I Kuramoto model Gr$(2,5)$ model or  $KM^I_{23}$.
  It contains the BS boundary St$^\mC_{32}\sim$U$(3)/$U$(1)\sim$SU$(3)$.
  The Kuramoto models on this configuration space were not considered before.
 Also  it would be interesting to analyzed
  the impact of the involutions described for the type II and III Kuramoto models.

We have not  carried out a study of the Kuramoto models related to
 the Cartan domain of type IV as well as for
the two exceptional domains of types V and VI. In particular,
the group action for the type IV domains is no longer the M\"{o}bius transform. For this
reason the Kuramoto equations should be different from the equations (\ref{f3}),(\ref{kii}) and (\ref{k3}).

We have not considered here  the limit $N\to\infty$. In this case the ensemble of oscillators
on the BS boundary is replaced on some distribution that evolves in time. The Poisson kernel
associated with the
corresponding domain allows one to construct special distributions on the domain itself. We plan to return
 to this problem later.


\appendix

\section{Boundeed symmetric domains}

\begin{center}
\begin{tabular}{|c|c|c|c|c|c|c|}
  \hline
  \hline

  Type &$G$ & U & $K$& $D$&$\dim\,D$\\
 \hline
 \hline
$ I_{mn}$ &$SU(m,n)$ & $SU(m+n)$ & $S(U(m)\times U(n))$ & $Z_{mn}$ & mn \\
 \hline
$II_{n}$ &$SO^*(2n)$ & $SO(2n)$   & $U(n)$ &$Z_{nn}=\{\bfz=-\bfz^T\}$&$n(n-1)/2$\\
\hline
$III_{n}$ &$Sp(n,\mR)$ & $Sp(n)$   & $U(n)$ &$Z_{nn}=\{\bfz=\bfz^T\}$&$n(n+1)/2$\\
\hline
$IV_{n}$ &$SO(n,2)$ & $SO(2+n)$   & $SO(n)\times SO(2)$ &$z\in\mC^n $&$n$\\
\hline
$V$ &$E_6(-14)$ & $E_6(-78)$   & $SO(10)\times SO(2)$ & $\clM_2$   &$16$\\
\hline
$VI$ &$E_7(-25)$ & $E_7(-133)$   & $E_6\times SO(2)$ &$\clM_3$&$27$\\
\hline
\hline
\end{tabular}

\bigskip

\end{center}
\noindent
Notation $E_6(-14)$ means that $-14$ is the character of the real form
of $E_6^\mC$. The similar notations we use for $E_7^\mC$ \cite{He}.\\
  $\bullet\, I_{mn}$ - the matrices $Z_{mn}$ satisfy inequality (\ref{ft}).\\
  $ \bullet\, II_{mn}$ - the matrices $Z_{mn}$ satisfy inequality (\ref{ii}).\\
 $\bullet\, III_{mn}$ - the matrices $Z_{mn}$ satisfy inequality (\ref{iii}).\\
 $\bullet\, IV_{mn}$ - the vector $z\in\mC^n$ satisfy inequalities $|z^Tz|^2-2z^\dag z+1>0$,
$|z^Tz|^2-1<0$.\\
$\bullet\, V$ $\clM_2$ - the matrices of order 2 over the Caley algebra
with operator norm less than 1.\\
$\bullet\, VI$ $\clM_3$ - the matrices of order 3 over the Caley algebra
with operator norm less than 1.\\

{\small {\bf Acknowledgments}\\
I would like to thank Dr P.Chekin for his helpful suggestions.
The work was supported by Russian Science Foundation
grant  24-12-00178.
} 


{\small



\end{document}